# Towards Provenance and Traceability in CRISTAL for HEP

**Jetendr Shamdasani**[1]**, Andrew Branson, Richard McClatchey**
University of the West of England
Coldharbour Lane, Bristol BS16 1, UK

E-mail: jetendr.shamdasani@cern.ch

**Abstract.** This paper discusses the CRISTAL object lifecycle management system and its use in provenance data management and the traceability of system events. This software was initially used to capture the construction and calibration of the CMS ECAL detector at CERN for later use by physicists in their data analysis. Some further uses of CRISTAL in different projects (CMS, neuGRID and N4U) are presented as examples of its flexible data model. From these examples, applications are drawn for the High Energy Physics domain and some initial ideas for its use in data preservation HEP are outlined in detail in this paper. Currently investigations are underway to gauge the feasibility of using the N4U Analysis Service or a derivative of it to address the requirements of data and analysis logging and provenance capture within the HEP long term data analysis environment.

## 1. Introduction

The analysis of High Energy Physics (HEP) data is often widely distributed among processing centres involving rapidly evolving suites of software and very complex datasets. One area that has been somewhat overlooked in HEP in recent years has been the tracking of HEP software development, its use in data analyses and its evolution over time. Tracking analyses in to provide historical records of the actions performed, the outcomes achieved and all (re-)design decisions taken is an important part of computer science research known as provenance data capture and management. Provenance management of scientific processes has become a very active area of research in the recent past. There has been a significant body of research that has been conducted in the computer science community on this topic; however, very little work has been done to address the specific provenance requirements that have emerged from the HEP domain in the Large Hadron Collider (LHC) era. This paper discusses a system known as CRISTAL which has been in development and active use at CERN for the past decade. CRISTAL is a mature and very stable system which was originally developed to track the construction of the Electromagnetic Calorimeter (ECAL) element of the Compact Muon Solenoid detector (CMS) [1] at CERN. The current usage is discussed in this paper in the context of its application at CMS and its use the neuGRID [2] and N4U (neuGRID for Users) [3] projects, where a so-called "Analysis Service" has been developed.

This paper is structured as follows: section 2 briefly discusses the CRISTAL system, section 3 describes its event based provenance model, and section 4 describes its application at the CMS ECAL. Section 5 discusses CRISTAL's use in the neuGRID and N4U projects. Section 6 presents some areas where we see further application in the HEP Community. Finally section 7 concludes and presents some areas for future work.

---

[1] Corresponding author

## 2. The CRISTAL System

CRISTAL is an object lifecycle oriented, description-driven platform for modelling processes and data structures via "descriptions", our term for model-as-data. It is completely traceable, meaning that any change made to the model or to the data instantiated from that model is logged together with the date of the change and who made the change, resulting in a complete history of changes (or the provenance of the item under consideration). All previous versions of the items are preserved and kept available for later use if required. The software was originally developed at CERN for tracking the assembly of the CMS ECAL. However, it has since evolved into a generic provenance capture and management and tool. CRISTAL is what is known as a *Description Driven System* (DDS). DDSs rely on configuration data to define their behaviour such that entire applications can be written without compiling additional code or interference with running systems. DDSs are used primarily for data-oriented applications; they provide powerful modelling techniques that represent a return to the original principles of object oriented design that modern scripting languages have moved away from.

The key CRISTAL concept is known as an *Item*. Items represent things or real world entities (similar to the notion of an Object in object oriented design). Within the item model, Items have their data structures defined as *XML schemas* and their contained application logic defined as *Scripts*. Every Item has a set of assigned *Activities* which are then assembled with control flow into *Lifecycles*. These can then be used to create new Items. Crucially, every object in the CRISTAL system is an Item, even the descriptions that define other descriptions. Briefly, Items in CRISTAL contain *Workflows*, that comprise Activities specifying work to be done by Agents (either human users or mechanical/ computational agents via an API), which then generate *Events* that detail each change of state of an Activity. Completion events generate data detailing the work done, known as *Outcomes* (which are XML documents from each execution), for which *Viewpoints* refer to particular historical versions (e.g. the latest version or, in the case of descriptions, a particular version number). *Properties* are name/value pairs that label and assign types to items, they also denormalize collected data for more efficient querying, and *Collections* which contain links to from one item to another. For a fuller explanation of the CRISTAL model and its application at CERN please see [4], and for an example of its practical use in neuroimaging analysis tracking see [5].

## 3. Provenance in CRISTAL

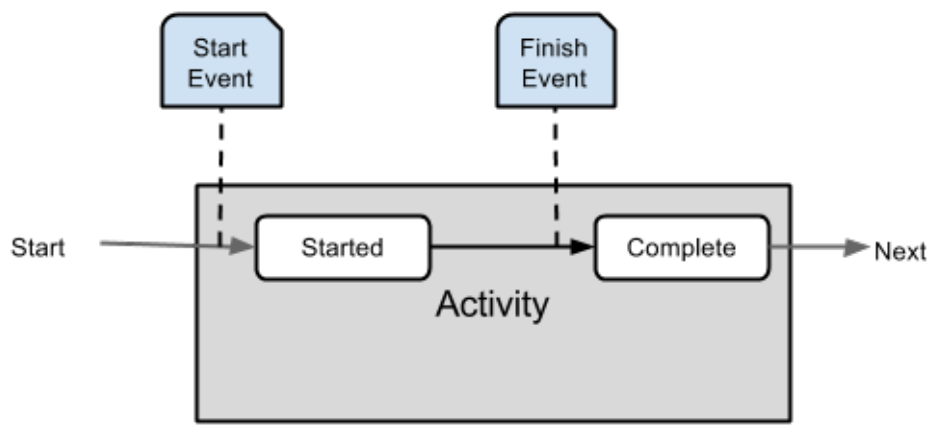

Figure 1: How and where Events are generated in an Activity

The key factor in CRISTAL, especially for the notion of provenance, is that when Items are created their older versions are not overwritten but are retained in a repository. Therefore, everything is kept and stored for later access if required. The key provenance related concept in CRISTAL is the notion of an Event. Every Activity has a state, and each interaction with it changes that state, usually progressing it towards completion. When an Activity changes state (for example when a request from an Agent starts or completes it) an Event is generated. This Event is an element of the provenance information that occurs when a state changes within an Activity. Figure 1 shows where in a CRISTAL Activity Event information is generated. In figure 1, an Event is generated before an Activity begins. This is to signify that the previous activity has completed (if there is one) or, if this is the first Activity in a Workflow, then it is to start the Workflow. Consequently once the Activity has started itself internally (such as a human operator begins filling a form) another Event is generated. Similarly for the end of an Activity, if it is an automatic Activity (such as a Script) then the end event is generated at the same time as the start.

The parameters that an Event contains are:
- *Event ID*: A unique ID for the Event.
- *Activity Name*: The name of the Activity that this Event is associated with.
- *Previous State*: This is the previous state of the activity before a change occurred.
- *Target State*: This is the state of the Activity after a change has occurred.
- *Transition*: This is the change that was initiated to the activity to cause the state of the activity to be altered.
- *Outcome Schema Name / Version*: This is the version name and number of the Outcome that is associated with this Event.
- *Agent Name*: The name of the Agent that executed the activity.
- *Agent Role*: The role of the Agent.
- *Time Stamp*: The time at which the Activity was started (for a start event) and finished (for a completion event).

When it was first designed the Event model in CRISTAL was created to capture the provenance of Workflows. As a consequence of the CRISTAL provenance model, Events can be linked and a graph can be created to show a detailed history of what has happened to an Activity. This information trace allows users to create an audit trail of what has happened in the past. This trace can be used for many different purposes, such as the re-creation of an experiment, or discovering what went wrong during the execution of a workflow. The next two sections outline examples of the use of a DDS by scientists in the construction of the CMS ECAL and in the neuGRID/N4U projects.

## 4. CRISTAL at the CMS ECAL

Originally proposed in 1997 and after a proof-of-concept prototype in 2000, CRISTAL was put into production in 2003 for tracking the assembly of the CMS ECAL Barrel, Endcap, and Preshower components across a set of distributed assembly centres. The production lifecycles of the components and sub-components changed many times during production, and our system seamlessly integrated these changes into a coherent set of data that has since been used for physics processes such as calibration and alignment. This was the first application of CRISTAL and was the reason the project was established. Here, it was used for schema, process and production model versioning.

There were three types of Items in the ECAL application: *Products*, which represented components of the detector which would be registered, characterized, then assigned to a position in the assembled detector structure; *Orders*, which would request a certain type of Product from a remote centre and specify the data structures required; and *Shipments*, which contained the packaged data relating to the physically delivered products. The Product lifecycle definitions, and the data schemas that related to them, evolved throughout the duration of the construction, and this was captured in CRISTAL.

## 5. CRISTAL in the NeuGRID and N4U Project

Since 2007, the University of the West of England (UWE) has contributed its software to the neuGRID consortium, applying CRISTAL to scientific analysis provenance; it is currently being used in the follow-on project N4U. In neuGRID it monitored grid workflow execution and constructed provenance data from this execution, therefore, it monitored *scientific workflow provenance*. In N4U, the system is being used to create what is known as the "Analysis Service". This is the full orchestration engine for executing analysis workflows, sending each job out individually to the Grid for processing. In comparison to neuGRID, it is currently being used in N4U to record the full *analysis provenance* for clinicians where its event model is being used to monitor the execution of image analysis workflows on the Grid. The system stores the result from every activity that is executed on the Grid and the parameters which were used to manipulate each image. Each data item at every step is recorded using CRISTAL's provenance model, and it is thereafter accessible at anytime by clinicians. The advantage of this is that it allows the *re-creation* of analyses by users, therefore, allowing the *re-producibility* of experiment

In addition to provenance management that is supported by CRISTAL, it has also been used for data integration, as a consequence of the flexibility of the description-driven nature of its underlying model. In N4U, it is used to facilitate the storage of any metadata that describes source data files alongside the files in a single repository. This repository is known as a "data atlas". This atlas of data is a large index of metadata which stores dynamic information on-the-fly for what users require in N4U. This is used by research clinicians to query and to discover datasets, subsets of datasets, different algorithms to run, analysis results and user supplied datasets. This large dataset of clinical information is a rich source of information for clinical researchers and provides them access to different pieces of data on their infrastructure by querying a single source where the metadata is stored. This is another promising application of CRISTAL as a dynamic database that can store any type of data for subsequent querying.

## 6. Future CRISTAL Application for Data Preservation in HEP

After the experience of using CRISTAL for provenance management in CMS and N4U we are currently considering its use for long-term data preservation in HEP (or DPHEP). There are many different scenarios and use cases where CRISTAL could be applied, and one tangible example is in the tracking of physics analyses. This builds on its previous application in the neuroimaging analysis domain and could bring concurrent algorithm versioning and evolution tracking, advanced traceability and solid reproducibility to HEP analysis effort. Another area where CRISTAL may be used in the HEP domains is as a flexible metadata index. As noted earlier in the N4U project, this was known as a "data atlas". Its unique input-validated (but schema-less) database design allows platform level data validation against schemas defined at runtime. This means that CRISTAL could store any XML data structure that has a schema. As a consequence this makes it ideal for indexing external metadata of file repositories. Allowing it to act as a giant queryable "data dump", this enables any form of XML-based data to be accessed at a later date and reused, when required, by physicists.

Open Data Publishing is another area where CRISTAL could be used for DPHEP. Within this area we are again able to harnesses the power of description-driven system flexibility, where all data within CRISTAL being queried as plain text (XML). As a result all analysis provenance data (and other information) can be made available to the public. Most data can be transformed to XML which although bulky, is extremely accessible compared to binary data formats. CRISTAL can publish data and schema this way, and log how that data is being used by communities of users. This inherent ability for simple and efficient storage and retrieval of data naturally leads to CRISTAL's application as a long term data archive. The same reasons that make XML so accessible make it an ideal long term archive format for data. Long after the detectors and analysis software is gone, the data and documentation describing it can be accessible to data archaeologists of the future. This allows all data to be preserved in a digital and human readable format for all future use.

## 7. Conclusions and Future Work

This paper has presented the CRISTAL software and its Item model has been presented briefly. For a full description the interested reader is directed to [4] and [5]. The CRISTAL provenance model was presented; it is focused on the provenance recording from Activities by generating Events which are recorded and stored. Some example uses were presented in different projects and ideas were given for its potential use in the DPHEP community. As well as CRISTAL's use in scientific research, it is currently being exploited commercially. It is currently commercialised by two companies. Agilium [6]

(Annecy, France), has developed a BPM (Business Process Management) solution and has been selling their product in the retail, logistics and manufacturing sectors since 2003. An EC FP7 IAPP project, CRISTAL-ISE [7], is increasing the knowledge transfer between Agilium and UWE through secondments, also involving a third company from the Manufacturing Execution Systems world, Alpha3i [8] of Rumilly, France. In another company, Technoledge [9] (Geneva, Switzerland), CRISTAL is being applied to fuel cell production lines with a focus on provenance data capture and management, therefore demonstrating its maturity as a provenance enabled system.

CRISTAL is currently available under a commercial license, but the next version (3.0) will be available under an open source license (LGPL). This is one of the goals of the CRISTAL-ISE project, and the codebase of CRISTAL is being overhauled and updated using techniques from the Semantic Web [10] and Distributed Systems [11] communities. One example of this modernization is making the CRISTAL provenance model more compliant with the Open Provenance Model (OPM) [12,13], allowing for interoperability with other provenance sources. Version 3.0 of the kernel component is currently planned for release mid 2014 and we are moving towards proposing this version to address the requirements of the DPHEP community.